# TUM

INSTITUT FÜR INFORMATIK

Concurrent Timed Port Automata

Radu Grosu
Bernhard Rumpe

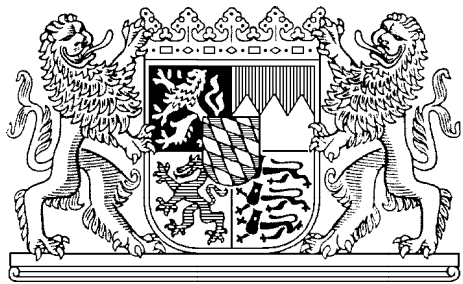

TECHNISCHE UNIVERSITÄT MÜNCHEN

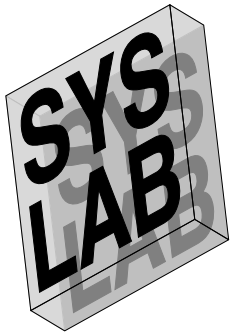

# Concurrent Timed Port Automata[*]

Radu Grosu, Bernhard Rumpe

Fakultät für Informatik, Technische Universität München
80290 München, Germany

E-Mail: {grosu,rumpe}@informatik.tu-muenchen.de

November 10, 2014

# Contents

**Abstract**

We present a new and powerful class of automata which are explicitly concurrent and allow a very simple definition of composition. The novelty of these automata is their time-synchronous message-asynchronous communication mechanism. Time synchrony is obtained by using a global clock. Message asynchrony is obtained by requiring the automata to react to every input. Explicit concurrency is obtained by marking each transition with a set of input and output messages. We compare these automata with a history based approach which uses the same communication mechanism and show that they are equivalent.

# Chapter 1

# Introduction

One of the most prominent models for interactive computation is the data-flow model. In this model, an interactive system is represented by a network of autonomous components which communicate solely by asynchronous transmission of messages through directed channels.

The components of a deterministic data-flow network can be characterized either in a history-based or in a state-based way. In the history-based approach [Kah74], each component is a function which incrementally reads from and writes onto its input and respectively output channels. In the state-based approach [Kok87, Jon89, LS89], each component is a deterministic automaton whose transitions are marked with input and output messages.

The extension of deterministic data-flow networks to nondeterministic data-flow networks proves, however, to be a nontrivial task. In the history-based approach, the replacement of functions with input/output relations leads to a non-compositional model, as shown in [BA81]. The replacement of functions with sets of functions, leads to a compositional model which is however too restrictive, because as shown in [Kel78], it cannot express nondeterministic fair merge components. In the state-based approach, there is no obvious way how to compose automata. One attempt is to take all possible synchronized interleavings, as done for I/O automata [LT89]. However, this departs from the original intuition and fails to model explicit concurrency.

The above difficulties are generated by an inappropriate level of abstraction. Nondeterministic models are useful only if they contain enough information about the causality between messages. Abstracting too much from this information not only destroys compositionality but also reduces the expressive power.

In [GS95] we proposed a *timed* model for the history-based approach where components are described with sets of functions. This model is compositional and fully abstract with respect to external observations. Moreover, it is powerful enough to describe not only nondeterministic fair merge components but also unbounded nondeterministic components.

Adding explicit time information to automata also leads to a very powerful, compositional model. This model is explicitly concurrent and very close to the original intuition. In this paper we present such a model and relate it to a slightly more general version of our history-based model. We show that they have equivalent expressive power. This also provides a full abstractness result for the history based model with respect to the more operational, state-based one.

The rest of the paper is split into five chapters. Chapter 2 introduces timed communication histories. Chapter 3 is concerned with the state-based approach: we define timed port automata and the composition of timed port automata. Chapter 4 is concerned with the history-based approach: we define components and the composition of components. In Chapter 5 we relate these two approaches and show their equivalence. Finally in Chapter 6 we discuss related work. To make our paper self contained we also provide an Appendix with basic notions about metric spaces.

# Chapter 2

# Basic Notions

We model an interactive system by a network of autonomous components which communicate via directed channels in a time-synchronous and message-asynchronous way. Time-synchrony is achieved by using a global clock splitting the time axis into discrete, equidistant time units. Message-asynchrony is achieved by allowing arbitrary, but finitely many messages to be sent along a channel in each time unit.

## 2.1 Communication Histories

We model the communication histories of directed channels by infinite streams of finite streams of messages. Each finite stream represents the communication history within a time unit. The first finite stream contains the messages received within the first time unit, the second the messages received within the second time unit, and so on (see Figure 2.1). Since time never halts, any complete communication history is infinite.

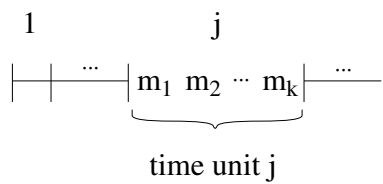

Figure 2.1: Timed stream

Let $D$ be the set of all messages. Then $[D^*]$ is the set of all complete communication histories, and $(D^*)^*$ is the set of all partial communication histories[1].

---

[1]For an arbitrary set $S$, we denote by $S^*$ the set of finite streams over $S$ and by $[S]$ the set of infinite streams over $S$.

Let $D$ be the set of all messages. Then $[D^*]$ is the set of all complete communication histories, and $(D^*)^*$ is the set of all partial communication histories[2].

A *named communication history* is a stream tuple $\iota \in (I \to [D^*])$ assigning to each channel named by the elements of $I$ a complete communication history[3]. Similarly, a *partial* named communication history $\iota \in (I \to (D^*)^*)$ assigns to each channel named by the elements of $I$ a partial communication history. Note that if $I = \emptyset$ then both $I \to [D^*]$ and $I \to (D^*)^*$ contain only one named stream tuple, the empty stream tuple and the empty partial stream tuple respectively.

Given a timed stream $s \in [D^*]$ and natural number $j$, $s\downarrow_j$ denotes the prefix of $s$ containing exactly the first $j$ sequences. The $\downarrow$ operator is overloaded to named stream tuples in a point-wise style, i.e. $\theta\downarrow_j$ denotes the result of applying $\downarrow_j$ to each component of $\theta$. In a point-wise style, it is also extended to sets of named stream-tuples.

### 2.1.1 Sum and Projection

Named sequences of messages and named communication histories can be combined and restricted by using the overloaded *sum* and *projection* operators. Before defining these operators let us denote by $\alpha\updownarrow_n \in I \to D^*$ the content of the named communication history $\alpha \in I \to [D^*]$ at time unit $n$.

**Definition 1 (Sum)** *For any $I, J$ such that $I \cap J = \emptyset$ we define the sum on named sequences as follows:*

$$+ \in (I \to D^*) \times (J \to D^*) \to ((I \cup J) \to D^*)$$
$$(\varphi + \psi)(i) = \begin{cases} \varphi(i) & \text{if } i \in I \\ \psi(i) & \text{if } i \in J \end{cases}$$

*We overload this sum to communication histories such that for all $n$:*

$$+ \in (I \to [D^*]) \times (J \to [D^*]) \to ((I \cup J) \to [D^*])$$
$$(\alpha + \beta)\updownarrow_n = \alpha\updownarrow_n + \beta\updownarrow_n \qquad \square$$

**Definition 2 (Projection)** *The projection of $\theta \in I \to D^*$ on $O$ is written as $\theta|_O$. It is an element of $(I \cap O) \to D^*$ such that:*

$$(\theta|_O)(i) = \theta(i) \quad \text{if } i \in (I \cap O)$$

*The projection of $\alpha \in I \to [D^*]$ on a set of names $O$ is also written as $\alpha|_O$. It is an element of $(I \cap O) \to [D^*]$ such that for all $n$:*

$$(\alpha|_O)\updownarrow_n = (\alpha\updownarrow_n)|_O \qquad \square$$

---

[2] For an arbitrary set $S$, we denote by $S^*$ the set of finite streams over $S$ and by $[S]$ the set of infinite streams over $S$.

[3] Note that $I \to [D^*]$ and $[I \to D^*]$ are isomorphic.

Note that if $I \cap O = \emptyset$ then $\theta|_O$ and $\alpha|_O$ represent the empty named sequence and the empty named stream tuple respectively. Projection is extended to sets of named stream-tuples in a point-wise style.

# Chapter 3

# The State-based Approach

## 3.1 Timed Port-Automata

In the state-based approach, we model data-flow components with timed port automata. Their interface consists of a set of input and output ports. To achieve modularity, output ports may be hidden.

**Definition 3 (Port signature)** *Let $D$ be a set of* data values *and $I, O, H$ pairwise disjoint sets of* input, output *and* hidden or internal ports *respectively. A port signature is a tuple $\Sigma = (D, I, O, H)$. Given a port signature $\Sigma = (D, I, O, H)$, we denote by $C = I \cup O \cup H$ the set of all ports in $\Sigma$.* □

As we already pointed out in the introduction, data-flow components communicate asynchronously. As a consequence, timed port automata are not allowed to block their environment. Therefore, in every state, they have to react to every possible input. Since the automata are timed, each reaction (or transition) takes place in time unit. The input or output associated to a transition may consist of a finite sequence of messages. The empty sequence denotes the absence of any message.

**Definition 4 (Timed port automaton)** *A timed port automaton is a tuple $A = (\Sigma, S, s^0, \delta)$ where:*
- *$\Sigma$ is a port signature,*
- *$S$ is a set of* states,
- *$s^0 \in S$ is the* start state,
- *$\delta \in S \times (C \to D^*) \times S$ is the transition relation, which is required to be* reactive:
  $$\forall s \in S, i \in (I \to D^*) : \exists t \in S, \theta \in (C \to D^*) : (s, \theta, t) \in \delta \land \theta|_I = i.$$ □

Note that reactiveness requires the existence of at least one transition in every state, even if the automaton does not have any input channels. This assures time progress because any transition takes place in one time unit. Another consequence is that a timed port automaton cannot have an empty transition relation.

If $(s, \theta, t) \in \delta$ we also write it as $s \xrightarrow{\theta}_\delta t$ or simply as $s \xrightarrow{\theta} t$ if $\delta$ is clear from the context. A named sequence $\theta \in N \to D^*$ is called an *input action* if $N = I$, an *output action* if $N = O$ and a *hidden action* if $N = H$. Input and output actions are also called *external actions*.

**Example 1 (The fair merge automaton)** *The automaton $FM = (\Sigma, \{s\}, s, \delta)$ where $\Sigma = (D, \{i, j\}, \{o\}, \emptyset)$, consumes data items from the channels $i$ and $j$ and produces them on $o$. The automaton is fair, i.e. it never neglects any incoming message indefinitely. The automaton has only one state which is also the initial one. The transition relation is defined as follows[1]:*

$$\delta = \{s \xrightarrow{\{i \mapsto a,\ j \mapsto b,\ o \mapsto c\}} s \mid \exists p \in \{i, j\}^{\#c} : a = pr_i(p, c) \land b = pr_j(p, c)\} \quad \text{where}$$
$$pr_k(k \ \&\ p, m \ \&\ a) = m \ \&\ pr_k(p, a)$$
$$pr_k(l \ \&\ p, m \ \&\ a) = pr_k(p, a)$$
$$pr_k(\epsilon, \epsilon) = \epsilon, \qquad \text{for } k, l \in \{i, j\} \text{ and } k \neq l$$

*Each transition corresponds to a merge of $a$ and $b$. Note that if $a$ or $b$ is $\epsilon$ then $c$ is $b$ or $a$ respectively.* □

## 3.2 Executions, Schedules and Behaviors

**Definition 5 (Execution, schedule, behavior)** *An* execution *of an automaton $A$ is an infinite sequence $s^0, \theta^0, s^1, \theta^1, \ldots$ such that $\forall i : s^i \xrightarrow{\theta^i} s^{i+1}$. We denote the set of executions by $execs(A)$.*

*The* schedule *$sched(\alpha)$ of an execution $\alpha$ is a subsequence of $\alpha$ containing only actions in $\alpha$. We denote the set of schedules by $scheds(A)$.*

*The* behavior *$beh(\alpha)$ of an execution or schedule $\alpha$ is the subsequence of $\alpha$ containing only external actions. We denote by $behs(\alpha)$ the set of all behaviors of $\alpha$.* □

Note that schedules and behaviors are named communication histories. Given an automaton $A$ and an input stream-tuple $\iota \in I \to [D^*]$. We denote the set of behaviors of $A$ with input $\iota$ by $A[\iota]$ and write it simply $[\iota]$ when $A$ is clear from the context. Formally:

$$A[\iota] = \{\alpha \in behs(A) \mid \alpha|_I = \iota\}$$

For a deterministic automaton $A[\iota]$ is a singleton for each $\iota$.

---

[1] $\#c$ is the length of the sequence $c$; $m \ \&\ a$ appends the message $m$ in front of the sequence $a$.

## 3.3 Strongly Pulse Driven Automata

Timed port automata have a very important property: *they process their input incrementally.* In other words, at any moment of time, their output does not depend on future input. This property is called *pulse-drivenness* and it has two variations: *strongly pulse-drivenness* and *weakly pulse-drivenness*.

The output produced in time unit $t$ by a strongly pulse-driven automaton, is not only independent from future input but also from input received in the same time unit. The output produced by a weakly pulse-driven automaton in time unit $t$ can also depend on the input received in time unit $t$. Hence, the first automaton introduces a delay between input and output while the second one may not.

**Definition 6 (Pulse driven automata)** *An automaton A is called* strongly pulse-driven *iff*

$$\forall \iota, \kappa, n : \ \iota \!\downarrow_n = \kappa \!\downarrow_n \ \Rightarrow \ [\iota]\,|_O\!\downarrow_{n+1} = [\kappa]\,|_O\!\downarrow_{n+1}$$

*An automaton A is called* weakly pulse-driven *iff*

$$\forall \iota, \kappa, n : \ \iota \!\downarrow_n = \kappa \!\downarrow_n \ \Rightarrow \ [\iota]\,|_O\!\downarrow_n = [\kappa]\,|_O\!\downarrow_n \qquad \square$$

Weakly pulse-drivenness can be defined equivalently as follows:

$$\forall \iota, \kappa, n : \ \iota \!\downarrow_n = \kappa \!\downarrow_n \ \Rightarrow \ [\iota] \!\downarrow_n = [\kappa] \!\downarrow_n$$

For a deterministic automaton, this property says that we can arrange the behaviors of the automaton in a tree with pairs $(\theta, \varphi) \in (I \to D^*) \times (O \to D^*)$ as nodes. The structure of the tree is given by the inputs, i.e., each node has $I \to D^*$ branches, one for each possible input. Each path in the tree corresponds to a behavior $\alpha$ and each level $n$ corresponds to the set of all behaviors $\alpha\!\downarrow_n$. The root has level 0. Behaviors with common input prefixes have a common sub-path to the root.

For a nondeterministic automaton, weakly pulse-drivenness says that, for every two inputs $\iota$ and $\kappa$ with common prefix of length $n$ and for every behavior $\alpha \in [\iota]$, we can find a behavior $\beta \in [\kappa]$ which has a common prefix of length $n$ with $\alpha$. Moreover, for a partial tree of behaviors and a new input $\iota$ we can always find a new behavior $\beta \in [\iota]$ which can be added to the tree. We first search in the tree for a behavior $\alpha$ which has the longest input prefix equal with $\iota$ and then choose from $[\iota]$ a $\beta$ which is equal on this prefix with $\alpha$.

**Theorem 1** *Every timed port automaton is weakly pulse-driven.*

**Proof:** For $\iota, \kappa \in I \to [D^*]$ suppose that

$$\iota\!\downarrow_n = \kappa\!\downarrow_n \quad \text{and} \quad [\iota]\,|_O\!\downarrow_n \neq [\kappa]\,|_O\!\downarrow_n$$

With no loss of generality, suppose that

$$\exists \gamma \in [\iota] : \ \gamma \downharpoonright_O \!\downarrow_n \notin [\kappa] \downharpoonright_O \!\downarrow_n.$$

Since $\gamma \in [\iota]$ it follows that $\gamma \downharpoonright_I \!\downarrow_n = \kappa \!\downarrow_n$. Let $\gamma_n = \gamma \!\downarrow_n$. By reactiveness of $A$ there is a $\delta$ such that

$$(\gamma_n \,\&\, \delta) \in behs(A) \quad \text{and} \quad (\gamma_n \,\&\, \delta)|_I = \kappa.$$

where & concatenates a finite stream with an infinite one. As a consequence $(\gamma_n \,\&\, \delta) \in [\kappa]$ which also implies that $\gamma \downharpoonright_O \!\downarrow_n \in [\kappa] \downharpoonright_O \!\downarrow_n$. This contradicts the assumption. □

**Theorem 2** *The fair merge automaton FM is weakly pulse-driven but not strongly pulse-driven.*

**Proof:** The output of $FM$ on the channel $o$ at time $t$ is completely determined by the input on the channels $i$ and $j$ at time $t$. □

Sometimes we want to require strong pulse-drivenness only on particular subsets of the input and output channels.

**Definition 7 (Strongly pulse-drivenness modulo $(J, P)$)** *A timed port automaton $A$ is strongly pulse-driven with respect to $(J, P)$, where $J \subseteq I$ and $P \subseteq O$, iff*

$$\forall \iota, \kappa, n : \ (\iota|_J) \!\downarrow_n = (\kappa|_J) \!\downarrow_n \ \wedge \ \iota|_{I \setminus J} = \kappa|_{I \setminus J} \ \Rightarrow \ ([\iota]|_P) \!\downarrow_{n+1} = ([\kappa]|_P) \!\downarrow_{n+1}$$

*Obviously, if $P = \emptyset$ then each automaton is strongly pulse-driven wrt. $(J, P)$.* □

An example of a strongly pulse-driven automaton is the buffer given below.

**Example 2 (A buffer with restricted delay)** *The buffer automaton*

$$BUF = (\Sigma, D^*, \epsilon, \delta) \quad \text{where} \quad \Sigma = (D, \{i\}, \{o\}, \emptyset)$$

*consumes data items from $i$ and reproduces them with a finite delay on $o$. The order of the incoming messages is preserved and the buffer capacity is unrestricted. The contents of the buffer are modeled by the set of states $D^*$. To ensure that every received message is also sent, we enforce $BUF$ to send at least one message if it is not empty. The transition relation is defined below, where & denotes the concatenation operation on finite sequences:*

$$\delta = \{a \,\&\, s \xrightarrow{\{i \mapsto b, \ o \mapsto a\}} s \,\&\, b \mid (a, b \in D^*) \wedge (a \,\&\, s \neq \epsilon \Rightarrow a \neq \epsilon)\}$$

*Note that the delay of every message is bounded by the former input.* □

## 3.4 One-to-Many Composition

The first thing we have to decide when composing automata is, if we allow them to share channels for writing. In the one-to-many composition only one automaton is allowed to write on a given channel. This assures that no interference (or merging) of messages can occur. We also require that hidden channels are private.

**Definition 8 (Compatible port signatures)** *The signatures $\Sigma_1$ and $\Sigma_2$ are called* compatible *iff $O_1 \cap O_2 = H_1 \cap C_2 = H_2 \cap C_1 = \emptyset$* □

If two automata are composed, the output channels of one automaton are connected to the input channels with the same name of the other automaton. A graphical illustration is given in Figure 3.1. The set of hidden channels of the composed automaton is the union of the sets of hidden channels of the components.

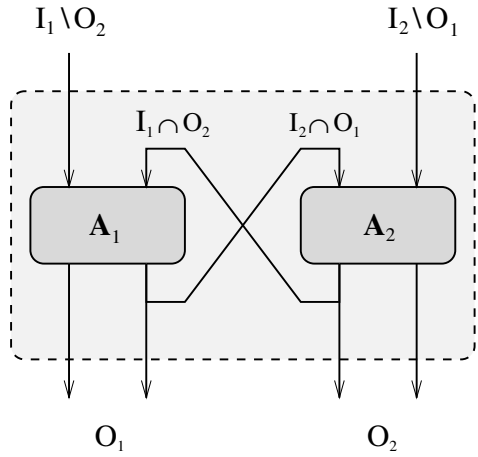

Figure 3.1: One-to-many composition

**Definition 9 (One-to-many composition of signatures)** *If $\Sigma_1$ and $\Sigma_2$ are compatible then the sets*

$$I = (I_1 \setminus O_2) \cup (I_2 \setminus O_1), \quad O = O_1 \cup O_2, \quad H = H_1 \cup H_2$$

*are pairwise disjoint. We may therefore define the* composition *of $\Sigma_1$ and $\Sigma_2$ to be the signature $\Sigma_1 \otimes \Sigma_2 = (I, O, H, D)$. As before $C = I \cup O \cup H$.* □

The main difficulty of untimed approaches is composition. The absence of any timing information about the causality or the relative timing of the messages exchanged by two automata leads to the Brock-Ackerman anomaly. The I/O automata approach introduces causality between messages by arranging them in a linear order: the synchronized merge of the messages of each automaton. Synchronization takes place when one automaton sends a message which the other automaton is waiting for. To make interleaving possible, each transition of an I/O automaton is marked with *only one* message: either an input or an output message. As a consequence, this approach is not explicitly concurrent. This is the case, even if one introduces special time transitions.

We work with a *global clock*: each transition takes place in the same, constant unit of time in both automata. Hence, our approach is *time-synchronous* and *message-asynchronous*. By using a global clock, we have enough causality information to obtain a very simple notion of composition: the set of input and output messages of the composed automaton at a given time $n$ is simply the union of the input and output messages of the components at the same time $n$. Note that union also takes care for communication. If an output channel of one automaton is connected to an input channel of the other automaton, the input channel does not appear in the composed automaton.

**Definition 10 (One-to-many composition of automata)** *Given two timed port automata $A_1$ and $A_2$ with compatible signatures. Then the* one-to-many composition *of $A_1$ and $A_2$ is $A_1 \otimes A_2 = (\Sigma, S, s^0, \delta)$ which is defined as follows:*

- $\Sigma = \Sigma_1 \otimes \Sigma_2$,
- $S = S_1 \times S_2$,
- $s^0 = (s_1^0, s_2^0)$,
- $\delta_1 \otimes \delta_2 \in S \times (C \to D^*) \times S$
  $\delta_1 \otimes \delta_2 = \{(s_1, s_2) \stackrel{\theta}{\to} (t_1, t_2) \mid \quad s_1 \stackrel{\theta|_{C_1}}{\to} t_1 \in \delta_1 \ \wedge \ s_2 \stackrel{\theta|_{C_2}}{\to} t_2 \in \delta_2\}$   □

Similarly to the composition for I/O automata, this composition allows automata to block each other, i.e., the composition can have an empty transition relation.

**Example 3 (Blocking)** *Given two automata with complementary signatures*

$A_1 = (\Sigma_1, \{s_1\}, s_1, \delta_1), \quad \Sigma_1 = (\mathbb{N}, \{i\}, \{o\}, \emptyset)$
$A_2 = (\Sigma_2, \{s_2\}, s_2, \delta_2), \quad \Sigma_2 = (\mathbb{N}, \{o\}, \{i\}, \emptyset)$

*and with the follwing transition relations, where $1 \,\&\, a$ appends the natural number 1 in front of the sequence $a$:*

$\delta_1 = \{(s_1, \{i \mapsto a, \ o \mapsto 1 \,\&\, a\}, s_1) \mid a \in \mathbb{N}^*\}$
$\delta_2 = \{(s_2, \{o \mapsto a, \ i \mapsto 1 \,\&\, a\}, s_2) \mid a \in \mathbb{N}^*\}$

*The composed automaton*

$A = (\Sigma_1 \otimes \Sigma_2, \{(s_1, s_2)\}, (s_1, s_2), \delta_1 \otimes \delta_2), \quad \Sigma_1 \otimes \Sigma_2 = (\mathbb{N}, \emptyset, \{i, o\}, \emptyset)$

*has an empty transition relation $\delta_1 \otimes \delta_2$, because there is no $\theta = \{i \mapsto a, \ o \mapsto b\}$ such that $b = 1 \,\&\, a$ as required by the first automaton and that $a = 1 \,\&\, b$ as required by the second automaton. This can be interpreted either as divergence or as blocking. In the second case each automaton waits without success for an output produced by the other automaton.*   □

Since $\delta_1 \otimes \delta_2$ is empty, $A$ is not a timed port automaton. This means that composition is a partial operation. However, if the output produced by one automaton is independent from the output produced by the other automaton then the automata cannot block each other and the composition is well defined.

**Theorem 3** *Given two timed port automata $A_1$ and $A_2$. Let $J = I_1 \cap O_2$ and $P = I_2 \cap O_1$. If either $A_1$ is strongly pulse-driven with respect to $(J, P)$ or $A_2$ is strongly pulse-driven with respect to $(P, J)$, then $\delta_1 \otimes \delta_2$ is well defined, that means $A_1 \otimes A_2$ is a port automaton.*

**Proof:** We have to show that $\delta_1 \otimes \delta_2$ is reactive. Given $s_1, s_2$ and $i \in I \to D^*$ and suppose that $A_1$ is strongly pulse-driven wrt. $(J, P)$ i.e its output on $P$ does not depend on its input on $J$.

This means, this output is completely determined by the state $s_1$ and the input $i$. Denote this output by $p$. Since $\delta_2$ is reactive, we can find for $s_2$ and $(i + p)|_{I_2}$ a $t_2$ and a $\theta_2$ such that $\theta_2|_{I_2} = (i + p)|_{I_2}$ and $(s_2, \theta_2, t_2) \in \delta_2$. Take $\theta|_{O_2} = \theta_2$.

This also fixes the input for $A_1$ on $J$ which together with $i$ determines a next state $t_1$ and an output $\theta_1$. Surely $\theta_1|_P = p$. Take $\theta|_{O_1} = \theta_1$.

Taking also $\theta|_I = i$ we obtain a transition $((s_1, s_2), \theta, (t_1, t_2)) \in \delta_1 \otimes \delta_2$. □

In the sequel we always assume that the conditions required by the above theorem hold. Note that if any of $J$ or $P$ is empty, the composition is well defined even if both automata are only weakly pulse-driven.

**Theorem 4** *Given timed port automata $A_1$ and $A_2$. Then*

$$execs(A_1 \otimes A_2) = \{e \mid e|_{A_1} \in execs(A_1) \wedge e|_{A_2} \in execs(A_2)\}$$
$$sheds(A_1 \otimes A_2) = \{e \mid e|_{A_1} \in sheds(A_1) \wedge e|_{A_2} \in sheds(A_2)\}$$
$$behs(A_1 \otimes A_2) = \{e \mid e|_{A_1} \in behs(A_1) \wedge e|_{A_2} \in behs(A_2)\}$$

**Proof:** Let

$$e = (s_1^0, s_2^0), \theta^0, (s_1^1, s_2^1), \theta^1, (s_1^2, s_2^2), \theta^2, \ldots$$

be an execution of $A_1 \otimes A_2$. Then

$$e|_{A_1} = s_1^0, \theta^0|_{C_1}, s_1^1, \theta^1|_{C_1}, s_1^2, \theta^2|_{C_1}, \ldots$$
$$e|_{A_2} = s_2^0, \theta^0|_{C_2}, s_2^1, \theta^2|_{C_2}, s_2^2, \theta^2|_{C_2}, \ldots$$

By definition of $\otimes$ and execution, for all $i$,

$$(s_1^i, s_2^i), \theta^i, (s_1^{i+1}, s_2^{i+1}) \quad \text{iff} \quad s_1^i, \theta^i|_{C_1}, s_1^{i+1} \quad \wedge \quad s_2^i, \theta^i|_{C_2}, s_2^{i+1}$$

As a consequence, $e|_{A_1} \in execs(A_1)$ and $e|_{A_2} \in execs(A_2)$. The proof for $sheds(A_1 \otimes A_2)$ and $behs(A_1 \otimes A_2)$ is similar. □

Strongly pulse-driveness is preserved by composition.

**Theorem 5** *If $A_1$ and $A_2$ are strongly pulse-driven automata then so is $A_1 \otimes A_2$.*
**Proof:** A straightforward consequence of Theorem 3. □

## 3.5 Hiding

Hiding is a very important operation which provides control over the scoping of channels. This is of great importance in the modular development of reactive systems.

There are two reasons for using a separate hiding operator instead of a composition which automatically hides the interconnected channels. First, the composition with hiding is not associative and commutative like the other one. Second, our compositionality result for behaviours could not have been formulated because the component behaviours would have contained information which were not present in the behaviour of the composed automata.

**Definition 11 (Hiding)** *Given a timed port automaton*

$$A = (\Sigma, S, s^0, \delta), \quad where \quad \Sigma = (D, I, O, H)$$

*and a set $P \subseteq O$. The automaton $\nu P : A$ is then defined as follows:*

$$\nu P : A = (\Sigma', S, s^0, \delta), \quad where \quad \Sigma' = (D, I, O \setminus P, H \cup P)$$

□

It is easy to see that $\nu P : A$ is a timed port automaton.

# Chapter 4

# The History-based Approach

## 4.1 Pulse-driven Functions

In the history-based approach we model data-flow components with sets of functions. Each function has the form $f \in (I \to [D^*]) \to (O \to [D^*])$. It maps named input histories to named output histories. The names of the input channels build the set $I$ and the names of the output channels build the set $O$.

The reason for working with infinite histories is that if no action is communicated along an input channel within a time unit, then an empty message sequence occurs in the input history informing the function that time has progressed. The lack of this timing information causes the fair merge anomaly [Kel78].

The functions should behave similar to deterministic automata i.e. they should process their input incrementally.

**Definition 12 (Pulse driven functions)** *Stream processing functions whose output until time $j$ $(j+1)$ is completely determined by the input until time $j$ are called weakly (strongly) pulse-driven. Formally:*

$$\forall \iota, \kappa, j : \iota\!\downarrow_j = \kappa\!\downarrow_j \Rightarrow f(\iota)\!\downarrow_j = f(\kappa)\!\downarrow_j \qquad \text{(weakly pulse-driven)}$$
$$\forall \iota, \kappa, j : \iota\!\downarrow_j = \kappa\!\downarrow_j \Rightarrow f(\iota)\!\downarrow_{j+1} = f(\kappa)\!\downarrow_{j+1} \quad \text{(strongly pulse-driven)}$$

*We use the arrow $\to$ for sets of strongly pulse-driven functions and the arrow $\xrightarrow{w}$ for sets of weakly pulse-driven functions.* □

Strongly and weakly pulse-driven functions have properties similar to deterministic automata. An appropriate framework to study these properties are metric spaces (see Appendix A and B).

**Theorem 6** *A stream processing function is strongly pulse-driven if and only if it is contractive with respect to the metric of streams. A stream processing function is* *weakly pulse-driven if and only if it is non-expansive with respect to the metric of streams.*

**Proof:** The metric of named stream tuples $(I \to [D^*]), d)$ with names in $I$ and elements in the discrete metric $(D, \rho)$ is defined as follows:

$$d(s,t) = inf\{2^{-j} \mid s\!\downarrow_j = t\!\downarrow_j\} \ .$$

This metric is complete [Eng77]. The contractive (non-expansive) functions $f$ satisfy $d(f(\iota), f(\kappa)) \leq c \cdot d(\iota, \kappa)$ with $c = 1/2$ ($c = 1$). $\square$

Thus, as a consequence of Banach's fixed point theorem, strong pulse-drivenness guarantees unique fixed points of feedback loops.

**Theorem 7** *Sum and projection are weakly pulse-driven functions.*

**Proof:** $(\iota + \kappa)\!\downarrow_n$ and $(\iota|_O)\!\downarrow_n$ only depend on $\iota\!\downarrow_n$ and $\kappa\!\downarrow_n$. $\square$

## 4.2 Components

We model static components by nonempty, closed sets of weakly pulse-driven functions.

**Definition 13 (Components)** *A static component whose input and output channels are named by $I$ and $O$, respectively, is modeled by a nonempty set of weakly pulse-driven functions*

$$F \subseteq (I \to [D^*]) \xrightarrow{w} (O \to [D^*]),$$

*that is* closed *in the sense that for all weakly pulse-driven functions $f$ of the same signature*

$$(\forall \iota \in (I \to [D^*]) : \exists f' \in F : f(\iota) = f'(\iota)) \Rightarrow f \in F. \qquad \square$$

The above definition is very powerful. It not only makes the model *fully abstract*, but it also allows to handle *unbounded nondeterminism*. Note that the use of relations instead of sets of functions is problematic in connection with unbounded nondeterminism (see for example [Cos85, NP92]).

## 4.3 One-to-many Composition

We define the composition slightly more general than in [GS95] because we relax the pulse-driveness requirements and allow one-to-many communication.

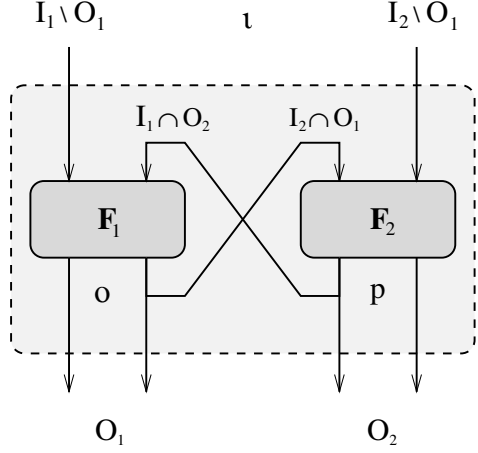

Figure 4.1: One-to-many composition

**Definition 14 (One-to-many composition)** *Given two components $F_1$ and $F_2$ with input channels in $I_1$ and $I_2$ and output channels in $O_1$ and $O_2$:*

$$F_1 \subseteq (I_1 \to [D^*]) \xrightarrow{w} (O_1 \to [D^*]), \qquad F_2 \subseteq (I_2 \to [D^*]) \xrightarrow{w} (O_2 \to [D^*])$$

*such that $I_1 \cap O_1 = I_2 \cap O_2 = O_1 \cap O_2 = \emptyset$. Let $J = I_1 \cap O_2$ and $P = I_2 \cap O_1$.*

*If either $F_1$ is strongly pulse-driven with respect to $(J, P)$ or $F_2$ is strongly pulse-driven with respect to $(P, J)$ then we may compose them in accordance with Figure 4.1. We refer to the resulting network as the* one-to-many *composition of $F_1$ and $F_2$ and write it as $F_1 \otimes F_2$. Formally:*

$$I = (I_1 \setminus O_2) \cup (I_2 \setminus O_1), \qquad O = O_1 \cup O_2,$$

$$F_1 \otimes F_2 \subseteq (I \to [D^*]) \xrightarrow{w} (O \to [D^*]),$$

$$F_1 \otimes F_2 = \{f \mid \forall \iota : \exists f_1 \in F_1, f_2 \in F_2 :$$
$$f(\iota) = o + p \text{ where } o = f_1((\iota + p)|_{I_1}), \ p = f_2((\iota + o)|_{I_2})\} \qquad \square$$

**Theorem 8** $F_1 \otimes F_2 \neq \emptyset$.

**Proof:** Since $F_1$ and $F_2$ are components we may find functions $f_1, f_2$ such that $f_1 \in F_1$ and $f_2 \in F_2$. The function:

$$g(p, \iota) = f_2((\iota + o)|_{I_2}) \quad \text{where} \quad o = f_1((\iota + p)|_{I_1})$$

where $o \in (O_1 \to [D^*])$ and $p \in (O_2 \to [D^*])$ is contractive in $p|_{I_1}$ as the composition of contractive and non-expansive functions. By Theorem 17 so is $\mu g$. Moreover, by contractiveness of $f_1$ and by non-expansiveness of Cartesian tuple projection, sum and stream tuple projection, $f$ defined by:

$$f(\iota) = o + p \quad \text{where} \quad o = f_1((\iota + p)|_{I_1}), \ p = (\mu g)(\iota)$$

is also non-expansive. Since for any predicate $P$:

$$(\exists f_1 \in F_1, f_2 \in F_2 : \forall \iota : P) \Rightarrow (\forall \iota : \exists f_1 \in F_1, f_2 \in F_2 : P),$$

we obtain $F_1 \otimes F_2 \neq \emptyset$. $\square$

The use of the universal quantification over input histories $\iota$ outside the existential quantification over functions $f_1 \in F_1, f_2 \in F_2$ assures the closeness of $F_1 \otimes F_2$.

**Theorem 9** $F_1 \otimes F_2$ *is a component.*

**Proof:** By Theorem 8 it is enough to prove that $F_1 \otimes F_2$ is closed. Suppose $f \in (I \to [D^*]) \xrightarrow{w} (O \to [D^*])$ and

$$\forall \iota : \exists f' \in F_1 \otimes F_2 : \ f(\iota) = f'(\iota).$$

Then for a given $\iota$ there is an $f' \in F_1 \otimes F_2$ an $f_1 \in F_1$ and an $f_2 \in F_2$ such that:

$$f'(\iota) = (o + p)|_O \ \ \text{where} \ \ o = f_1((\iota + p)|_{I_1}), \ \ p = f_2((\iota + o)|_{I_2}).$$

Hence, for every $\iota$ there are functions $f_1 \in F_1$ and $f_2 \in F_2$ such that:

$$f(\iota) = (o + p)|_O \ \ \text{where} \ \ o = f_1((\iota + p)|_{I_1}), \ \ p = f_2((\iota + o)|_{I_2}).$$

In other words $f \in F_1 \otimes F_2$. $\square$

## 4.4 Hiding

As for timed port automata, hiding provides control over the scoping of channels. This is of great importance in the modular development of reactive systems.

**Definition 15 (Hiding)** *Given a component $F$ with input channels in $I$ and output channels in $O$*

$$F \subseteq (I \to [D^*]) \xrightarrow{w} (O \to [D^*])$$

*and a set of channel names $P \subseteq O$. The component $\nu P : F$ is defined as follows:*

$$\nu P : F \subseteq (I \to [D^*]) \xrightarrow{w} ((O \setminus P) \to [D^*])$$
$$\nu P : F = \{f \ | \ \forall \iota : \exists g \in F : \ f(\iota) = g(\iota)|_{O \setminus P}\}$$

$\square$

**Theorem 10** *Given a component $F$ with inputs in $I$ and outputs in $O$ and a set $P \subseteq O$. Then $\nu P : F$ is also a component.*

**Proof:** Trivial. $\square$

# Chapter 5

# Equivalence of Approaches

Given a timed port automaton $A$ we can generate a component $[\![A]\!]$ by associating to each tree of behaviors of the automaton a weakly pulse-driven function.

**Definition 16 (The function set generated by an automaton)** *Given a timed port automaton $A$ with signature $\Sigma = (I, O, H, D)$. Then we can associate a set $[\![A]\!]$ to $A$ as follows:*

$$[\![A]\!] = \{ f \in (I \to [D^*]) \xrightarrow{w} (O \to [D^*]) \mid$$
$$\forall \iota \in (I \to [D^*]) : \exists \alpha \in behs(A) : \iota = \alpha|_I \wedge f(\iota) = \alpha|_O \} \qquad \Box$$

The generation process has some important properties.

**Theorem 11** *Given a timed port automaton $A$. Then for each $\eta \in behs(A)$ there is a function $f \in [\![A]\!]$ such that $f(\eta|_I) = \eta|_O$.*
**Proof:** In the following we show that for each behavior $\eta \in behs(A)$, there is a tree-like set $S$ of behaviors $\alpha \in behs(A)$ which contains $\eta$ and corresponds to a weak pulse-driven function.

We call a set $S$ of behaviours $\alpha \in behs(A)$ weakly pulse-driven (or tree-like) if:

$$\forall \alpha, \beta \in S, n \in \mathbb{N} : (\alpha|_I)\downarrow_n = (\beta|_I)\downarrow_n \quad \Rightarrow \quad \alpha\downarrow_n = \beta\downarrow_n$$

$S$ characterizes a function $f$ only partially because it does not necessarily contain a behavior for each possible input. $S$ is called *maximal* if

$$\forall \iota \in (I \to [D^*]) : \iota \in S|_I$$

If the set $S$ is not maximal, then it can be enlarged by adding a new behavior. Let $\iota \notin S|_I$ be an input.

If there is a behavior $\alpha \in S$ such that $\alpha|_I$ has the greatest common prefix with $\iota$, whose length is $n$, then the weak pulse drivenness of $A$ guarantees the existence of a behavior $\beta \in A[\iota]$, such that $\beta\downarrow_n = \alpha\downarrow_n$.

If there is no such $\alpha$, then $\iota$ must be the limit of a Cauchy sequence $\{\alpha_n|_I \mid n \in \mathbb{N}\}$ with $\alpha_n$ satisfying

$$\forall n : \exists \alpha_n \in S : (\alpha|_I)\!\downarrow_n = \iota\!\downarrow_n$$

Since $S$ is weakly pulse-driven, the set $\{\alpha_n \mid n \in \mathbb{N}\}$ also builds a Cauchy sequence. The upper limit $\beta$ of this sequence is by construction in $A[\iota]$.

In both cases, the set $S \cup \{\beta\}$ is again weakly pulse-driven and $\beta \notin S$. Given a chain $S_i$ of weakly pulse-driven sets, the union $S = \bigcup_i S_i$ is itself weakly pulse-driven, because for all $\alpha, \beta \in S$, there is already a chain element $S_i$ with $\alpha, \beta \in S_i$. By Zorn's lemma, we get maximal elements in the cpo of weakly pulse-driven sets ordered by set inclusion. These elements are exactly our maximal weakly pulse-driven sets. Maximal sets are usually uncountable, and therefore not constructive, but Zorn ensures their existence.

Now starting with $\{\eta\}$ we enlarge it to a maximal weakly pulse-driven set $S$. Then we define $f$ by

$$\forall \iota : \exists \alpha \in S : \alpha|_I = \iota \wedge f(\iota) = \alpha|_O.$$

Maximality of $S$ ensures uniqueness of $f$. Weak pulse-drivenness of $S$ ensures well definedness and weak pulse-drivenness of $f$. Moreover, the construction of $S$ also ensures that $f(\eta|_I) = \eta|_O$ and $f \in [\![A]\!]$. $\square$

**Theorem 12** *The set $[\![A]\!]$ associated to a timed port automaton $A$ is a component.*

**Proof:** First, we have to show that $[\![A]\!] \neq \emptyset$. By definition of $A$, $\delta \neq \emptyset$, so the set of behaviors $behs(A) \neq \emptyset$. By theorem 11, there is at least one weakly pulse driven function for each behavior. Hence $[\![A]\!] \neq \emptyset$.

Second, we have to show $[\![A]\!]$ is closed. Suppose

$$f \in (I \to [D^*]) \xrightarrow{w} (O \to [D^*])$$

and

$$\forall \iota : \exists f' \in [\![A]\!] : f(\iota) = f'(\iota).$$

This means, each behavior of $f$ is in $behs(A)$. Hence, $f$ is also in $[\![A]\!]$. $\square$

Now we are ready to state the main result of this chapter. Given two timed port automata $A_1$ and $A_2$, the one-to-many composition $[\![A_1]\!] \otimes [\![A_2]\!]$ of the components generated by $A_1$ and $A_2$ is the same as the component $[\![A_1 \otimes A_2]\!]$ generated by the one-to-many composition $A_1 \otimes A_2$ of the automata $A_1$ and $A_2$.

**Theorem 13** $[\![A_1 \otimes A_2]\!] = [\![A_1]\!] \otimes [\![A_2]\!]$.

**Proof:**

$\subseteq$: Suppose that $f \in [\![A_1 \otimes A_2]\!]$. Then by definition of $[\![.]\!]$

$$\forall \iota \in (I \to [D^*]) : \exists \alpha \in behs(A_1 \otimes A_2) : \iota = \alpha|_I \wedge f(\iota) = \alpha|_O.$$

By definition of $.\otimes.$

$$\forall \alpha \in behs(A): \ \alpha|_{A_1} \in behs(A_1) \ \wedge \ \alpha|_{A_2} \in behs(A_2).$$

Hence, by theorem 11, there must exist $f_1 \in [A_1]$ and $f_2 \in [A_2]$ such that

$$f_1(\alpha|_{I_1}) = \alpha|_{O_1}, \quad f_2(\alpha|_{I_2}) = \alpha|_{O_2}$$

Observing that

$$\alpha|_O = \alpha|_{O_1} + \alpha|_{O_2}, \quad \alpha|_{I_1} = (\iota + \alpha|_{O_2})|_{I_1}, \quad \alpha|_{I_2} = (\iota + \alpha|_{O_1})|_{I_2}$$

we obtain

$$\forall \iota \in (I \to [D^*]) : \exists f_1 \in [A_1], f_2 \in [A_2]: \ f(\iota) = o + p \quad \text{where}$$
$$o = f_1((\iota + p)|_{I_1}), \quad p = f_2((\iota + o)|_{I_2})$$

Hence $f \in [A_1] \otimes [A_2]$.

$\supseteq$: Suppose $f \in [A_1] \otimes [A_2]$. Then by definition of $.\otimes.$

$$\forall \iota \in (I \to [D^*]) : \exists f_1 \in [A_1], f_2 \in [A_2]: \ f(\iota) = o + p \quad \text{where}$$
$$o = f_1((\iota + p)|_{I_1}), \quad p = f_2((\iota + o)|_{I_2})$$

Let $\alpha = \iota + o + p$. Then, by definition of $\alpha$

$$\alpha|_{A_1} = (\iota + p)|_{I_1} + o, \quad \alpha|_{A_2} = (\iota + o)|_{I_2} + p.$$

Since $f_1 \in [A_1]$ and $f_2 \in [A_2]$ it is the case that

$$\alpha|_{A_1} \in behs(A_1), \quad \alpha|_{A_2} \in behs(A_2).$$

Hence $\alpha \in behs(A_1 \otimes A_2)$. Because $f(\alpha|_I) = \alpha|_O$ we obtain that $f \in [A_1 \otimes A_2]$. $\square$

# Chapter 6

# Related Work

The theory of automata is a very prolific field of computer science. A thorough survey of the work in this field could be alone the subject of a paper. In the following we only mention those which influenced our own work.

Similarly to the I/O automata [LT89], our automata use a message-asynchronous communication mechanism. The input enableness condition of the former is closely related to our reactiveness condition. However, the transitions of I/O automata are marked with only one message and their composition is defined differently. Timed I/O automata seem to be more closely related to ours [LV91]. However, their composition remains in principle the same as for I/O automata and their timing concept is more powerful but also more complicated than ours.

Similarly to the automata informally described in [Kok87], our automata have parallel interfaces and work in a timed environment. However composition of timed automata is not studied in [Kok87], because the main emphasis of this work is to give a fully abstract denotational semantics for untimed data-flow networks. Similar timed automata were also used in [Jon89]. These are however composed by taking all possible synchronized interleavings.

The relational theory of nondeterministic data-flow networks is also the subject of intensive research.

The model presented in this paper is a slight variation of the one given [GS95]. In that model we used point-to-point communication and and required the functions to be strongly pulse-driven on all channels, whereas in this model we use one-to-many communication, an additional hiding operator and require the functions to be strongly pulse-driven only on the feedback channels.

For both models we had several sources of inspiration. First of all, they are inspired by [Par83], [Kok87], [Bro87]. Park models components by sets of functions in the same way as we do. However, he models time with time ticks $\sqrt{}$ and his functions are defined also for finite streams. Moreover, infinite streams are not required to have infinitely many ticks. Kok models nondeterministic components by functions mapping timed streams to sets of timed streams. We use instead a closed set of deterministic pulse-driven functions. This allows us to model unbounded nondeterminism without having to introduce a more complex metric. [Bro87] employs sets of functions as we do, but these functions work on untimed finite and infinite streams. This makes the model more abstract but at the same time more complex with respect to its theoretical basis. The formulation of pulse-drivenness has been taken from [Bro95], and the use of named communication histories is based on [BD92]. The use of closed sets of functions to model dataflow components is not new — see for example [Rus90].

# Acknowledgments

This work could not have been possible without the valuable support of Manfred Broy who also read an earlier version of this paper. Thanks go also to Ketil Stølen for stimulating discussions and valuable feedback and to Peter Scholz for reading an earlier version of this paper. The authors have been financially supported by the DFG project SYSLAB.

# Appendix A

# Metric Space Definitions

## A.1 Metric Space Basics

The fundamental concept in metric spaces is the concept of distance.

**Definition 17 (Metric Space)** *A* metric space *is a pair $(D, d)$ consisting of a nonempty set $D$ and a mapping $d \in D \times D \to \mathbb{R}$, called a* metric *or* distance, *which has the following properties:*

- **(1)** $\forall x, y \in D: \quad d(x, y) = 0 \quad \Leftrightarrow \quad x = y$
- **(2)** $\forall x, y \in D: \quad d(x, y) = d(y, x)$
- **(3)** $\forall x, y, z \in D: \quad d(x, y) \leq d(x, z) + d(z, y)$ □

A very simple example of a metric is the discrete metric.

**Definition 18 (The discrete metric)** *The* discrete metric $(D, d)$ *over a set $D$ is defined as follows:*

$$d(x, y) = \begin{cases} 0 & \text{if } x = y \\ 1 & \text{if } x \neq y \end{cases}$$
□

Measuring the distance between the elements of a sequence $(x_i)_{i \in \mathbb{N}}$ in $D$ we obtain the familiar definitions for convergence and limits.

**Definition 19 (Convergence and limits)** *Let $(D, d)$ be a metric space and let $(x_i)_{i \in \mathbb{N}}$ be a sequence in $D$.*

**(1)** *We say that $(x_i)_{i \in \mathbb{N}}$ is a* Cauchy sequence *whenever we have:*

$$\forall \epsilon > 0: \exists N \in \mathbb{N}: \forall n, m > N: d(x_n, x_m) < \epsilon.$$

**(2)** *We say that* $(x_i)_{i \in \mathbb{N}}$ *converges to* $x \in D$ *denoted by* $x = lim_{n \to \infty} x_i$ *and call* $x$ *the* limit *of* $(x_i)_{i \in \mathbb{N}}$ *whenever we have:*

$\forall \epsilon > 0 : \exists N \in \mathbb{N} : \forall n > N : d(x_n, x) < \epsilon.$

**(3)** *The metric space* $(D, d)$ *is called* complete *whenever each Cauchy sequence converges to an element of* $D$. □

**Theorem 14** *The discrete metric is complete.*
**Proof:** Each Cauchy sequence is constant from a given $N$. □

A very important class of functions over metric spaces is the class of *Lipschitz functions*.

**Definition 20 (Lipschitz functions)** *Let* $(D_1, d_1)$ *and* $(D_2, d_2)$ *be metric spaces and let* $f \in D_1 \to D_2$ *be a function. We call* $f$ Lipschitz function *with constant* $c$ *if there is a constant* $c \geq 0$ *such that the following condition is satisfied:*

$d(f(x), f(y)) \leq c \cdot d(x, y)$

*For a function* $f$ *with arity* $n$ *the above condition generalizes to:*

$d(f(x_1, \ldots, x_n), f(y_1, \ldots, y_n)) \leq c \cdot max\{d(x_i, y_i) \mid i \in [1..n]\}$

*If* $c = 1$ *we call* $f$ non-expansive. *If* $c < 1$ *we call* $f$ contractive. □

**Theorem 15** *The composition of two Lipschitz functions* $f \in D_1 \to D_2$ *and* $g \in D_2 \to D_3$ *is a Lipschitz function with constant* $c_1 \cdot c_2$.
**Proof:** $d(g(f(x_1)), g(f(x_2))) \leq c_2 \cdot d(f(x_1), f(x_2)) \leq c_2 \cdot c_1 \cdot d(x_1, x_2)$ □

**Lemma 1** *The composition of a contractive and a non-expansive function is contractive. The composition of two non-expansive functions is non-expansive. Identity is non-expansive.* □

The main tool for handling recursion in metric spaces is the Banach's fixed point theorem. It guarantees the existence of a unique fixed point for every contractive function.

**Theorem 16 (Banach's fixed point theorem)** *Let* $(D, d)$ *be a complete metric space and* $f \in D \to D$ *a contractive function. Then there exists an* $x \in D$, *such that the following holds:*

**(1)** $x = f(x)$      ($x$ *is a fixed point of* $f$)
**(2)** $\forall y \in D : y = f(y) \Rightarrow y = x$    ($x$ *is unique*)
**(3)** $\forall z \in D : x = lim_{n \to \infty} f^n(z)$    *where*
     $f^0(z) = z$
     $f^{n+1}(z) = f(f^n(z))$

**Proof:** See [Eng77] or [Sut75]. □

Usually we want to use a parameterized version of this theorem.

**Definition 21 (Parameterized fixed point)** *Let $f \in D \times D_1 \times \ldots \times D_n \to D$ be a function of non-empty complete metric spaces that is contractive in its first argument. We define the* parameterized fixed point *function $\mu f$ as follows:*

$(\mu f) \in D_1 \times \ldots \times D_n \to D$
$(\mu f)(y_1, \ldots, y_n) = x$

*where $x$ is the unique element of $D$ such that $x = f(x, y_1, \ldots, y_n)$ as guaranteed by Banach's fixed point theorem.* □

**Theorem 17** *If $f$ is contractive (non-expansive) so is $\mu f$.*

**Proof:** See for example [MPS86] pages 114–115. □

## A.2  Streams and Named Stream Tuples

A stream is a finite or infinite sequence of elements. For any set of elements $E$, we use $E^*$ to denote the set of all finite streams over $E$, and $[E]$ to denote the set of all infinite streams over $E$. For any infinite stream $s$, we use $s\!\downarrow_j$ to denote the prefix of $s$ containing exactly $j$ elements. We use $\epsilon$ to denote the empty stream.

We define the metric of streams generically with respect to an arbitrary discrete metric $(E, \rho)$.

**Definition 22 (The metric space of streams)** *The metric space of streams $([E], d)$ over a discrete metric $(E, \rho)$ is defined as follows:*

$[E] = \prod_{i \in \mathbb{N}} E$
$d(s, t) = inf\{2^{-j} \mid s\!\downarrow_j = t\!\downarrow_j\}$

□

This metric is also known as the Baire metric [Eng77].

**Theorem 18** *The metric space of streams $([E], d)$ is complete.*

**Proof:** See for example [Eng77]. □

A *named stream tuple* is a mapping $\theta \in (I \to [E])$ from a set of names to infinite streams. $\downarrow$ is overloaded to named stream tuples in a point-wise style, i.e. $\theta\!\downarrow_j$ denotes the result of applying $\downarrow_j$ to each component of $\theta$.

**Definition 23 (The metric of named stream tuples)** *The metric of* named *stream tuples* $(I \to [E], d)$ *with names in* $I$ *and elements in* $(E, \rho)$ *is defined as follows:*

$I \to [E]$ *is the set of functions from the countable set* $I$ *to the metric* $[E]$,

$d(s,t) = inf\{2^{-j} \mid s\!\downarrow_j = t\!\downarrow_j\}$

□

**Theorem 19** *The metric space of named stream tuples* $(I \to [E], d)$ *is complete.*

**Proof:** This metric is equivalent to the Cartesian product metric $\prod_{i \in I}[E]$ which is complete because $[E]$ is [Eng77]. □